\documentclass[10pt,twocolumn,twoside]{IEEEtran}

\usepackage[mathscr]{eucal}
\usepackage[tight,footnotesize]{subfigure}
\usepackage[cmex10]{amsmath}
\usepackage{cite}
\usepackage{afterpage,layout,longtable,varioref,verbatim}
\usepackage{algorithmic}
\usepackage{array}
\usepackage{graphics}
\usepackage{color}
\usepackage{latexsym}
\usepackage{epsf,psfrag}
\usepackage{epsfig}
\usepackage{amssymb}
\usepackage{textcomp}
\usepackage{amsmath}
\usepackage{bm}

\newtheorem{theorem}{Theorem}[section]

\newtheorem{definition}{Definition}[section]

\newcommand{\eg}{\textit{e.g.}}
\newcommand{\ie}{\textit{i.e.}}

\newcommand{\etal}{\textit{et al.}}

\definecolor{dkr}{rgb}{0.6,0.2,0.2}

\definecolor{light-gray}{gray}{0.85}
\definecolor{BrickRed}{RGB}{150,25,14}

\newcommand{\Amsc}{\mathscr{A}}

\newcommand{\Emsc}{\mathscr{E}}

\newcommand{\Gmsc}{\mathscr{G}}

\newcommand{\Nmsc}{\mathscr{N}}

\newcommand{\Rmsc}{\mathscr{R}}
\newcommand{\Smsc}{\mathscr{S}}
\newcommand{\SmscA}{\mathscr{S}_{\mbox{\scriptsize A}}}
\newcommand{\SmscF}{\mathscr{S}_{\mbox{\scriptsize F}}}

\newcommand{\Vmsc}{\mathscr{V}}

\newcommand{\mbbE}{\mathbb{E}}  
\newcommand{\mbbR}{\mathbb{R}}  

\newcommand{\Nc}{\mathcal{N}}

\newcommand{\Bbf}{{\bf {B}}}
\newcommand{\Hbf}{{\bf {H}}}

\newcommand{\Ibf}{{\bf {I}}}
\newcommand{\Pbf}{{\bf {P}}}
\newcommand{\Qbf}{{\bf {Q}}}
\newcommand{\Rbf}{{\bf {R}}}

\newcommand{\Wbf}{{\bf {W}}}

\newcommand{\abf}{{\bf {a}}}

\newcommand{\ebf}{{\bf {e}}}

\newcommand{\rbf}{{\bf {r}}}

\newcommand{\qbf}{{\bf {q}}}
\newcommand{\sbf}{{\bf {s}}}

\newcommand{\xbf}{{\bf {x}}}
\newcommand{\ybf}{{\bf {y}}}
\newcommand{\zbf}{{\bf {z}}}

\newcommand{\Fscr}{\mbox{\scriptsize F}}

\newcommand{\lisp}{\vspace{5pt}}
\newcommand{\slisp}{\vspace{3pt}}

\def\QEDclosed{\mbox{\rule[0pt]{1.3ex}{1.3ex}}}
\def\QED{\QEDclosed}
\def\endproof{\hspace*{\fill}~\QED\par\endtrivlist\unskip}

\newcommand{\bitem}{\begin{itemize}}
\newcommand{\eitem}{\end{itemize}}
\newcommand{\bearr}{\begin{equation*}\begin{array}}
\newcommand{\eearr}{\end{array}\end{equation*}}
\newcommand{\beq}{\begin{equation*}}
\newcommand{\eeq}{\end{equation*}}
\newcommand{\bea}{\begin{eqnarray}}
\newcommand{\eea}{\end{eqnarray}}

\begin{document}

\title{Data Framing Attack on State Estimation\thanks{J. Kim, L. Tong, and R. J. Thomas are with the School of Electrical and Computer Engineering, Cornell University, Ithaca, NY 14853, USA. Email: {\tt \{jk752, ltong, rjt1\}@cornell.edu}.  Part of this work was presented at the Asilomar Conference on Signals, Systems, and Computers, Pacific Grove, CA, November, 2013.
}}

\author{Jinsub Kim,
        Lang Tong,~\IEEEmembership{Fellow,~IEEE},
        and~Robert J. Thomas,~\IEEEmembership{Life Fellow,~IEEE}}%


\maketitle{\let\thefootnote\relax\footnotetext{This work is supported in part by the NSF under Grant CNS-1135844 and the Army Research Office under Grant W911NF1010419.}}

\begin{abstract}
A new mechanism aimed at misleading a power system control center about the source of a data attack is proposed.  As a man-in-the-middle state attack, a data framing attack is proposed to exploit the bad data detection and identification mechanisms currently in use at most control centers.  In particular, the proposed attack frames meters that are providing correct data as sources of bad data such that the control center will remove useful measurements that would otherwise be used by the state estimator.  

The optimal design of a data framing attack is formulated as a quadratically constrained quadratic program (QCQP). It is shown that the proposed attack is capable of perturbing the power system state estimate by an arbitrary degree controlling only half of a critical set of measurements that are needed to make a system unobservable.  Implications of this attack on power system operations are discussed, and the attack performance is evaluated using benchmark systems.

\end{abstract}

\begin{IEEEkeywords}
Power system state estimation, bad data test, data framing attack, cyber security, smart grid.
\end{IEEEkeywords}

\section{Introduction}
\IEEEPARstart{A}{} feature of any future smart grid is the promise of a data-driven approach to automated monitoring, control, and decision as opposed to the current simulation-driven methods.  The paradigm shift to a data-driven framework enables a deeper integration of data collection and sophisticated data processing into the monitoring and control process.  While extracting actionable information from real-time sensor data can make a grid more efficient and adaptive to real-time operating conditions, it exposes the grid to possible cyber data attacks aimed at disrupting grid operations and potentially causing blackouts.

In \cite{Liu:2009CCS}, Liu, Ning, and Reiter presented perhaps the first framework for a man-in-the-middle (MiM) attack on the power system state estimation where an adversary would replace ``normal'' sensor data with ``malicious data.''
It was shown that, if the adversary could gain control of a sufficient number of meters, it could perturb the state estimate by an arbitrary amount without being detected by the bad data detector employed at the control center.  Such undetectable attacks are referred to as {\em covert data attacks}.

The condition under which a covert data attack is possible was subsequently found in \cite{Kosut11} to be equivalent to that of system unobservability.  In particular, a covert attack is possible if and only if the system becomes unobservable when the meters under attack are removed.  (or equivalently, the adversary is able to control a critical set of meters.) 
The minimum number of meters that an adversary has to control in order to launch a covert data attack, referred to as a {\em security index}, is an important measure of security against a data attack.  It  represents a fundamental limit on the capability of an adversary to covertly disrupt the operation of a grid \cite{Sandberg&Teixerira&Johansson:10SCS, Kosut11}.

In this paper, we show that a significant barrier on the capability of an adversary to mount an attack of the type described above can be circumvented by using a different form of attacks, that is, one that exploits the vulnerabilities of the existing bad data detection and removal mechanisms.  In particular, we show that the adversary only needs to gain control of about half of the meters required by the security index while achieving the same objective of perturbing the state estimate by an arbitrary amount without being detected.  

The attacks considered in this paper are referred to as {\em data framing attacks}, borrowing the notion of framing as that of providing false evidence to make someone innocent appear to be guilty of misconduct.  In the context of state estimation, a data framing attack means that an adversary launches a data attack in such a way that the control center identifies properly functioning meters as sources of bad data.  To this end, the attacker does not try to cause malicious data to pass the bad data detection without detection (as a covert attack tries to do).   Instead, it purposely triggers the bad data detection mechanism and causes erroneous removal of good data.  Unknown to the control center, the remaining data still contain adversary-injected malicious data, causing errors in the state estimate.

\subsection{Related work}
There is an extensive literature on {\em covert data attacks}, following the work of Liu, Ning, and Reiter \cite{Liu:2009CCS}.  While the data framing attack mechanism proposed here is fundamentally different, insights gained from existing work are particularly relevant.  Here, we highlight some of these ideas from the literature.

An explicit link between a covert attack on state estimation and system observability was made in  \cite{Kosut&Jia&Thomas&Tong:10SGC,Bobba&etal:10SCS}.  Consequently, classical observability conditions \cite{Krumpholz&Clements&Davis:80PAS,Monticelli&Wu:85TPAS,Monticelli&Wu:85TPASb} can be modified for a covert attack and used to  develop meter protection strategies \cite{Bobba&etal:10SCS, Kosut11, Kim&Poor:2011TSG, Bi&Zhang:2011Globecom, GianiEtal:2011SGC, KimTong:13JSAC}.  A particularly important concept is the notion of a critical set of meters \cite{Krumpholz&Clements&Davis:80PAS,Clements&Davis:1986TPD,Korres&Contaxis:91TPS}.
In assessing the vulnerability of the grid, the minimum number of meters necessary for a covert attack was suggested as a security index for the grid in \cite{Sandberg&Teixerira&Johansson:10SCS, Kosut11}.
Subsequently, meter protection strategies were proposed in \cite{Dan&Sandberg:10SGC, Vukovic2011} to optimize this security index.  

The framing attack strategy considered here relies on bad data identification and removal techniques that have long been the subject of study \cite{Handschin&Schweppe&Kohlas&Feichter:75TPAS, Monticelli&Garcia:1983TPAS, VanCutsem85TPAS, Clements&Davis:1986TPD,Mili&VanCustem&Ribbens-Pavella:1984TPAS}.  See for example \cite{Abur&Exposito:book, Monticelli:book} and the references therein. Typically,  the residue vectors in normalized forms are widely used as statistics for the bad data test \cite{Handschin&Schweppe&Kohlas&Feichter:75TPAS}. 
 In particular, Mili \etal \cite{Mili&VanCustem&Ribbens-Pavella:1984TPAS} proposed a hypothesis testing method, in which the set of suspect measurements are determined by the residue analysis in \cite{Handschin&Schweppe&Kohlas&Feichter:75TPAS}. 
The use of non-quadratic cost functions in state estimation was also studied to enhance bad data identification performance.  Especially, the weighted least absolute value estimation \cite{KotiugaVidyasagar:1982TPS, AburCelik:91TPS, CelikAbur:92TPS, SinghAlvarado:1994TPS} and the least median of squares regression \cite{MiliEtal:94TCS, CheniaeEtal:96TPS} were considered as alternatives with 
comparably good performance. 
In this paper, we take the residue analysis in \cite{Handschin&Schweppe&Kohlas&Feichter:75TPAS} as a representative bad data test and analyze the effect of a framing attack.  However, the same analysis is applicable to other bad data tests.

Detection of data attacks on state estimation, referred to as \emph{state attacks}, has been also studied in various frameworks.
Kosut~\etal\cite{Kosut11} presented a generalized likelihood ratio test for detection.
Morrow~\etal\cite{MorrowEtal:2011HICSS} proposed the detection mechanism based on network parameter perturbation which deliberately modifies the line parameters and probes whether the measurements respond accordingly to the modification.  
Distributed detection and estimation of adversarial perturbation was also studied in \cite{TajerEtal:2011SGC}.
In an effort to minimize the detection delay, the attack detection was also formulated as a quickest detection problem, and modified CUSUM algorithms were proposed in \cite{HuangEtal:2011CISS, CuiEtal:2012SPM, HuangEtal:2013CommMag}.

\subsection{Summary of results and organization}
We propose a data framing  attack on power system state estimation.  Specifically, we formulate the design of an optimal data framing attack as a quadratically constrained quadratic program (QCQP).  
Unlike general QCQPs, which are NP-hard, the proposed QCQP can be solved by finding a maximum eigenvalue of a matrix.  
To analyze the efficacy of our data framing attack, we present a sufficient condition under which the attack could achieve an arbitrary perturbation of the state estimate by controlling only half of the critical set of meters.
We demonstrate the concept using both the IEEE 14-bus network and the IEEE 118-bus network and show that the sufficient condition holds for the critical sets associated with cuts.

The optimal design of our framing attack is based on a linearized system.  In practice, a nonlinear state estimator is often used.
We demonstrate that, under the usual nonlinear measurement model, a framing attack designed based on a linearized system model  successfully perturbs the state estimate, and the attacker is able to control the degree of perturbation as desired.

The remainder of the paper is organized as follows: 
Section~\ref{sec:prelim_baddata} introduces the measurement and adversary models including preliminaries related to state attacks.  Section~\ref{sec:SE} presents the state estimation and bad data processing methodology.
In Section~\ref{sec:frame_attack}, we present the main idea of the data framing attack and the QCQP framework for the attack design. Section~\ref{sec:factor_of_two} provides a theoretical justification of the efficacy of the data framing attack.
In Section~\ref{sec:numerical_baddata}, we test the data framing attack with the IEEE 14-bus network and the IEEE 118-bus network.
Finally, Section~\ref{sec:conclusion_frame} provides concluding remarks.

\section{Mathematical models}\label{sec:prelim_baddata}

This section introduces the topology and system state of a power network, the meter measurement model, and the adversary model.  In addition, the covert state attack and its connection with network observability are explained.
Throughout the paper, boldface lower case letters $(\eg, \xbf)$ denote vectors, $x_{i}$ denotes the $i$th entry of the vector $\xbf$, boldface upper case letters $(\eg, \Hbf)$ denote matrices, $\Hbf_{ij}$ denotes the $(i,j)$ entry of $\Hbf$, $\Rmsc(\Hbf)$ denotes the column space of $\Hbf$, $\Nmsc(\Hbf)$ denotes the null space of $\Hbf$, and script letters $(\eg, \Smsc, \Amsc)$ denote sets.  The multivariate normal distribution with the mean $\bm{\mu}$ and the covariance matrix $\bm{\Sigma}$ is denoted by $\Nc(\bm{\mu}, \bm{\Sigma})$.

\subsection{Network and measurement models}

A power network is a network of buses connected by transmission lines, and thus the \emph{topology} of the grid can be naturally defined as an undirected graph $\Gmsc = (\Vmsc, \Emsc)$ where $\Vmsc$ is the set of buses, and $\Emsc$ is the set of lines connecting buses ($\{i, j\}\in\Emsc$ if and only if bus $i$ and bus $j$ are connected.)
The \emph{system state} of a power network is defined as the vector of bus voltage magnitudes and phase angles, from which all the other quantities (\eg, power line flows, power injections, line currents) can be calculated.

In order to compute a real-time estimate of the system state, a control center collects measurements from line flow and bus injection meters\footnote{Other types of meters can also be considered, but we restrict our attention to line flow and bus injection meters for simplicity.} deployed throughout the grid.  
 The meter measurements are related to the system state $\xbf$ in a nonlinear fashion, and the relation is described by the AC power flow model\cite{Abur&Exposito:book}:
\begin{equation}\label{eq:AC_baddata}
\zbf = h(\xbf) + \ebf,
\end{equation}
where $h(\cdot)$ is the nonlinear measurement function, and $\ebf$ is the Gaussian measurement noise with a diagonal covariance matrix.

If some of the meters malfunction or an adversary injects malicious data, the control center observes biased measurements,
\begin{equation}
\bar{\zbf} = h(\xbf) + \ebf + \abf,
\end{equation}
 where $\abf$ represents a deterministic bias.  In such a case, the data are said to be \emph{bad}, and the biased meter entries are referred to as \emph{bad data entries}.
Note that even when a meter is protected from adversarial modification, it may still have a bias due to a physical malfunction or an improper parameter setting; filtering out the measurements from such malfunctioning meters was the original objective of the legacy bad data processing and is still in practice today \cite{Handschin&Schweppe&Kohlas&Feichter:75TPAS}.

Even though the model in (\ref{eq:AC_baddata}) is nonlinear, the state estimate is generally obtained by iterations of weighted linear least squares estimation with the locally linearized model \cite{Abur&Exposito:book}.  
Therefore, it is reasonable to analyze the performance of state estimation using the locally linearized model around the system operating point.
To this end, in analyzing the impact of an attack on state estimation, we adopt the so-called DC model \cite{Abur&Exposito:book}.
In the DC model, for ease of analysis, the AC model (\ref{eq:AC_baddata}) is linearized around the system state where all voltage phasors are equal to $1\angle 0$, and only the real part of the measurements are retained:
\begin{equation}\label{eq:DC_baddata}
\zbf = \Hbf\xbf + \ebf,
\end{equation}
where $\zbf\in\mbbR^{m}$ is the measurement vector consisting of real part of line flow and bus injection measurements, the system state $\xbf \in \mbbR^{n}$ is the vector of voltage phase angles at all buses except the reference bus ($\xbf$ is unknown, but deterministic), $\Hbf\in\mbbR^{m\times n}$ is the DC measurement matrix that relates the system state to bus injection and line flow amounts, and $\ebf$ is the Gaussian measurement noise with a diagonal covariance matrix $\bm{\Sigma}$.  We represent the noise covariance matrix $\bm{\Sigma}$ as $\bm{\Sigma} = \sigma^{2}\bar{\bm{\Sigma}}$, where $\bar{\bm{\Sigma}}$ is a diagonal matrix representing the variation of noise variances across different meters ($\sum_{i=1}^{m}\bar{\bm{\Sigma}}_{ii} = 1$), and $\sigma^{2}$ is a scaling factor.

Each row of $\Hbf$ has a special structure depending on the type of the meter \cite{Abur&Exposito:book}.  For ease of presentation, consider the noiseless measurement $\zbf = \Hbf\xbf$.  If an entry $z_{k}$ of $\zbf$ is the measurement of the line flow from bus $i$ to bus $j$, $z_{k}$ is $B_{ij}(x_{i} - x_{j})$ where $B_{ij}$ is the line susceptance and $x_{i}$ is the voltage phase angle at bus $i$ \cite{Abur&Exposito:book}.  If $z_{k}$ is the measurement of bus injection at $i$, it is the sum of all the outgoing line flows from $i$, and the corresponding row of $\Hbf$ is the sum of the row vectors corresponding to all the outgoing line flows.

Any analysis based on the DC model needs to be verified using realistic AC model simulations; we demonstrate in Section~\ref{sec:numerical_baddata} that the proposed attack strategy is effective using AC model simulations.

\subsection{An adversary model}

We consider a man-in-the-middle attack on power system state estimation, where, as described in Fig.~\ref{fig:se_attack_analog}, an adversary is assumed to be capable of modifying the data from a subset of analog meters $\SmscA$.  We refer to the meters in $\SmscA$ as \emph{adversary meters}.

The control center observes corrupted measurements $\bar{\zbf}$ instead of the actual measurements $\zbf$ in (\ref{eq:AC_baddata}).  We assume that the adversary knows the line parameters (\ie, the measurement function $h$ and the measurement matrix $\Hbf$.)  

The adversarial modification is mathematically modeled as follows:
\begin{equation}\label{eq:attack_baddata}
\bar{\zbf} = \zbf + \abf,~~~\abf\in\Amsc,
\end{equation}
where $\abf$ is an attack vector, and $\Amsc$ is the set of feasible attack vectors defined as 
\begin{equation}
\Amsc\triangleq\{\abf\in\mbbR^{m}:~a_{i} = 0,~\forall i\notin\SmscA\}.
\end{equation}
Note that $\Amsc$ fully characterizes the ability of the adversary. 
In addition, the adversary is assumed to design a vector $\abf$ without observing any entry of $\zbf$, \ie, the attack does not require any real-time observation.

\subsection{Network observability and covert state attacks}\label{subsec:observability}

For state estimation to be feasible, the control center needs to have enough meter measurements so that the system state can be uniquely determined.
Formally, a power network is said to be \emph{locally observable} at a state $\xbf_{0}$ if the system state can be uniquely determined from the noiseless meter measurements $h(\xbf)$ in a neighborhood of $\xbf_{0}$.  This implies that the Jacobian of $h$ at $\xbf_{0}$ has full rank.  
However, due to the intractability of checking local observability for all feasible operating points,
the DC model (\ref{eq:DC_baddata}) is generally adopted for observability analysis \cite{Krumpholz&Clements&Davis:80PAS}: the network is said to be \emph{observable} if the DC measurement matrix $\Hbf$ has full rank.
In practice, power networks should be designed to satisfy the observability requirement.  Hence, we assume that the network of interest is observable (\ie, $\Hbf$ has full rank.)


The concept of network observability is closely related to the feasibility of a covert state attack.
In particular, we need to introduce the concept of a critical set of meters, formally defined as follows.  
\lisp
\begin{definition}\label{def:critical}
A subset $\Smsc_{0}$ of meters is said to be a \emph{critical set} if removing all meters in $\Smsc_{0}$ from the network makes the network unobservable whereas removing any strict subset of $\Smsc_{0}$ does not.
\end{definition}
\lisp

The covert state attack was first proposed in \cite{Liu:2009CCS} for the DC model, and it is formally defined as follows.
\lisp
\begin{definition}
Given a measurement vector $\zbf = \Hbf\xbf + \ebf$, an attack $\abf$ is said to be \emph{covert} if $\bar{\zbf}$ is equal to $\Hbf\bar{\xbf} + \ebf$ for some $\bar{\xbf}\neq \xbf$.
\end{definition}
\lisp

Note that if measurements are perturbed by a covert attack, the corrupt measurements appear to be normal measurements from the state $\bar{\xbf}$.  
From the above definition, an attack $\abf$ is covert if and only if $\abf$ is equal to $\Hbf\ybf$ for some nonzero $\ybf\in\mbbR^{n}$.  
In \cite{Kosut11}, the condition for existence of a covert attack was characterized as a network unobservability condition, as stated in the following theorem.
\lisp
\begin{theorem}[\hspace{-0.2pt}\cite{Kosut11}]\label{thm:Kosut}
A covert attack exists if and only if removing the adversary meters renders the network unobservable (or equivalently, the attacker can control at least a critical set of meters.)  
In addition, if $\abf$ is covert, then so is $\gamma\cdot\abf$, and $\|\bar{\xbf} - \xbf\|_{2}$ increases to infinity as $\gamma$ grows.  
\end{theorem}

\section{State estimation and bad data processing}\label{sec:SE}

This section introduces a popular approach to state estimation and bad data processing, which we assume to be employed by the control center.  
Once the control center receives measurements $\zbf$, it aims to obtain an estimate $\hat{\xbf}$ of the system state $\xbf$.  Because bad data entries in $\zbf$ may result in a bias in the state estimate, the control center employs a mechanism to filter out possible bad data entries in $\zbf$.  

\begin{figure}[t!]
\centering
\psfrag{z}[c]{ $\zbf$ }
\psfrag{zb}[c]{ $\bar{\zbf}$ }
\psfrag{g}[c]{ $\Gmsc$}
\psfrag{gb}[c]{ ${\Gmsc}$}
\psfrag{x}[c]{ $\hat{\xbf}$ }
\psfrag{o}[l]{ $(\hat{\xbf},\Gmsc)$}
\psfrag{ob}[l]{$\hat{\xbf}$}
\psfrag{n}[c]{ $\sbf$}
\includegraphics[width=.48\textwidth]{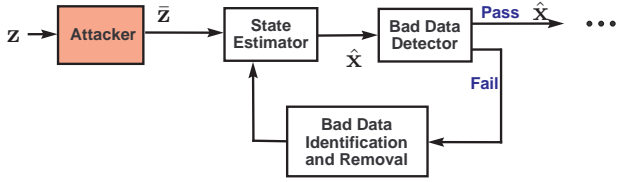}
\caption{Adversary model with state estimation and bad data test}
\label{fig:se_attack_analog}
\end{figure}

Fig.~\ref{fig:se_attack_analog} illustrates an iterative scheme for obtaining $\hat{\xbf}$, which consists of three functional blocks: state estimation, bad data detection, and bad data identification \cite{Handschin&Schweppe&Kohlas&Feichter:75TPAS, Abur&Exposito:book}.  
The iteration begins with the initial measurement vector $\zbf^{(1)} \triangleq \zbf$ and the initial measurement function $h^{(1)} \triangleq h$ where the superscript denotes the index for the current iteration.  

\begin{table}[t!]
\centering
\begin{small}
\caption{Iterative state estimation}\label{table:pseudocode}
\fbox{\parbox{3.2in}
{ $\text{Iterative-State-Estimation}(\zbf, h, \bm{\Sigma})$\vspace{.5em}
	

1:\hspace{1.0em} $\zbf^{(1)} \leftarrow \zbf$; $~~$ $h^{(1)} \leftarrow h$; $~~$ $k \leftarrow 1$; \vspace{.5em}

2:\hspace{1.0em} while ($true$)\vspace{.5em}

3:\hspace{2.0em} $(\hat{\xbf}^{(k)}, \rbf^{(k)}) \leftarrow \text{State-Estimation}(\zbf^{(k)}, h^{(k)})$;\vspace{.5em}

4:\hspace{2.0em} $result \leftarrow \text{Bad-Data-Detection}(\rbf^{(k)})$;\vspace{.5em}

5:\hspace{2.0em} if $result == good$\vspace{.5em}

6:\hspace{3.0em} break;\vspace{.5em}

7:\hspace{2.0em} else \vspace{.5em}

8:\hspace{3.0em} $(\zbf^{(k+1)}, h^{(k+1)}) \leftarrow \text{Bad-Data-ID}(\rbf^{(k)}, \zbf^{(k)}, h^{(k)})$;\vspace{.5em}

9:\hspace{2.0em} end \vspace{.5em}

10:\hspace{1.5em} $k \leftarrow k+1$; \vspace{.5em}

11:\hspace{0.5em} end \vspace{.5em}

12:\hspace{0.5em} return $\hat{\xbf}^{(k)}$; \vspace{.5em}

}}
\end{small}
\end{table}

\subsection{State estimation and bad data detection}

At the $k$th iteration, the state estimator uses $(\zbf^{(k)}, h^{(k)})$ as an input, and obtains the weighted least squares (WLS) estimate of the system state:
\begin{equation}\label{eq:WLS_AC}
\hat{\xbf}^{(k)} \triangleq \arg\min_{\xbf}(\zbf^{(k)} - h^{(k)}(\xbf))^{T}(\bm{\Sigma}^{(k)})^{-1}(\zbf^{(k)} - h^{(k)}(\xbf)),
\end{equation}
where $\bm{\Sigma}^{(k)}$ is the covariance matrix of the corresponding noise vector. 
Based on the state estimate, the residue vector is also evaluated: 
\begin{equation}
\rbf^{(k)} \triangleq \zbf^{(k)} - h^{(k)}(\hat{\xbf}^{(k)}).
\end{equation}


We assume that the $J(\hat{\xbf})$-test\cite{Handschin&Schweppe&Kohlas&Feichter:75TPAS, Abur&Exposito:book} is employed for bad data detection: the bad data detector makes a decision based on the sum of weighted squared residues: 
\begin{equation}\label{eq:BDD}
\left\{\begin{array}{ll}
\text{bad data} & \text{if $(\rbf^{(k)})^{T}(\bm{\Sigma}^{(k)})^{-1}\rbf^{(k)} > \tau^{(k)}$;}\\
\text{good data} & \text{if $(\rbf^{(k)})^{T}(\bm{\Sigma}^{(k)})^{-1}\rbf^{(k)} \leq \tau^{(k)}$.}
\end{array}
\right.
\end{equation}
The $J(\hat{\xbf})$-test is widely used due to its low complexity and the fact that the test statistic has a $\chi^{2}$ distribution if the data are good \cite{Handschin&Schweppe&Kohlas&Feichter:75TPAS}.  The latter fact is used to set the threshold $\tau^{(k)}$ for a given false alarm constraint.  


\subsection{Iterative bad data identification and removal}

If the bad data detector (\ref{eq:BDD}) declares that data are good, the algorithm returns the state estimate $\hat{\xbf}^{(k)}$ and terminates.  
However, if the bad data detector declares that the data are bad, bad data identification is invoked to identify and remove one bad data entry from the measurement vector.

A widely used criterion for identifying a bad data entry is the normalized residue\cite{Handschin&Schweppe&Kohlas&Feichter:75TPAS, Abur&Exposito:book}, which is considered one of the most reliable criteria\cite{VanCutsem85TPAS}.  
In the normalized residue analysis, each $r^{(k)}_{i}$ is divided by its standard deviation under the good data hypothesis (\ie, the standard deviation of $r^{(k)}_{i}$ when there exists no bad data entry in $\zbf^{(k)}$.)
If there exists no bad data entry in $\zbf^{(k)}$, and the state estimate $\hat{\xbf}^{(k)}$ is close to the actual state $\xbf$, the distribution of $\rbf^{(k)}$ can be approximated by $\Nc(\textbf{0}, \Wbf^{(k)}\bm{\Sigma}^{(k)})$ where 
\begin{equation}
\Wbf^{(k)} \triangleq \Ibf - \Hbf^{(k)}((\Hbf^{(k)})^{T}(\bm{\Sigma}^{(k)})^{-1}(\Hbf^{(k)}))^{-1}{(\Hbf^{(k)})}^{T}(\bm{\Sigma}^{(k)})^{-1}
\end{equation}
 with $\Hbf^{(k)}$ denoting the Jacobian of $h^{(k)}$ at $\hat{\xbf}^{(k)}$ and $\Ibf$ denoting the identity matrix with the appropriate size (see Appendix of \cite{Handschin&Schweppe&Kohlas&Feichter:75TPAS} for the detail.)  
Hence, the normalized residue is calculated as
\begin{equation}
\tilde{\rbf}^{(k)} = \bm{\Omega}^{(k)}\rbf^{(k)},
\end{equation}
where $\bm{\Omega}^{(k)}$ is a diagonal matrix with 
\begin{equation}\label{eq:norm_matrix}
\bm{\Omega}^{(k)}_{ii} =\left\{\begin{array}{ll}
 0 & \text{if $\{i\}$ is a critical set\footnotemark,}\\
 \frac{1}{\sqrt{(\Wbf^{(k)}\bm{\Sigma}^{(k)})_{ii}}} & \text{otherwise.}
\end{array}\right.
\end{equation}
\footnotetext{If $\{i\}$ is a critical set (\ie, removing the meter $i$ makes the grid unobservable), its residue is always equal to zero\cite{Abur&Exposito:book}, and the corresponding diagonal entry of $\Wbf^{(k)}\bm{\Sigma}^{(k)}$ is zero.  For such a meter, the normalizing factor is $0$ such that its normalized residue is equal to $0$.}

Once the normalized residue $\tilde{\rbf}^{(k)}$ is calculated, the meter with the largest $|\tilde{r}^{(k)}_{i}|$ is identified as a bad meter.  
The bad data identification unit removes the row of $\zbf^{(k)}$ and the row of $h^{(k)}$ that correspond to the bad meter and returns the updated measurement vector and measurement function for the next iteration, denoted by $\zbf^{(k+1)}$ and $h^{(k+1)}$.

Table~\ref{table:pseudocode} provides the pseudocode for the overall procedure of the iterative state estimation and bad data test.

Using the DC model (\ref{eq:DC_baddata}), state estimation, bad data detection, and bad data identification are the same with that in the AC model, except that the nonlinear measurement function $h^{(k)}(\xbf)$ is replaced with the linear function $\Hbf^{(k)}\xbf$ (so, the Jacobian is the same everywhere.) 
Note that the WLS state estimate (\ref{eq:WLS_AC}) is replaced with a simple linear WLS solution:
\begin{equation}\label{eq:WLS_DC}
\hat{\xbf}^{(k)} = ((\Hbf^{(k)})^{T}(\bm{\Sigma}^{(k)})^{-1}(\Hbf^{(k)}))^{-1}{(\Hbf^{(k)})}^{T}(\bm{\Sigma}^{(k)})^{-1}\zbf^{(k)},
\end{equation}
and thus
\begin{equation}\label{eq:r_DC}
\rbf^{(k)} = \zbf^{(k)} - \Hbf^{(k)}\hat{\xbf}^{(k)} = \Wbf^{(k)}\zbf^{(k)}.
\end{equation}

\section{Data framing attack}\label{sec:frame_attack}

In this section, we present a new attack strategy on state estimation, referred to as a \emph{data framing attack}, which exploits the bad data processing to remove data from some normally operating meters and make the adversary meters appear to be trustworthy. 
We present the main idea and the QCQP framework for an optimal attack strategy.  

We focus our attention on the case where the adversary cannot control enough meters to launch a covert attack. 
A framing attack starts with setting the set of \emph{framed meters}, denoted by $\SmscF$, which it aims to frame as sources of bad data.  
The framed meter set $\SmscF$ is chosen such that after the meters in $\SmscF$ are removed from the network, a covert attack with the adversary meters in $\SmscA$ becomes feasible.
For instance, suppose that $\Smsc$ is a critical set. 
 The feasibility condition of a covert attack in Theorem~\ref{thm:Kosut} implies that if $\Smsc\setminus\SmscA$ is removed from the network, then the adversary with $\SmscA$ can launch a covert attack; because, further removing all meters in $\SmscA$ makes the network unobservable.  Therefore, a framing attack can set $\SmscF$ to be $\Smsc\setminus\SmscA$.  

The resulting perturbation of the state estimate by a framing attack does depend on the choice of $\SmscF$. 
Finding an optimal $\SmscF$ for a given attack objective is certainly an important problem.
However, it is out of scope of this paper.  We focus on the design of the attack vector for a fixed $\SmscF$.


\subsection{Effect of attack on normalized residues}\label{subsec:effect_rN}
To analyze how an attack affects the bad data processing, we analyze, using the DC model (\ref{eq:DC_baddata}), the adversarial effect on the normalized residue vector at the first iteration.  
In this subsection, we omit the superscript to simplify notation: all the quantities we consider are associated with the first iteration unless otherwise specified.

Suppose that $\zbf$ is a measurement vector \emph{without} bad data.  The normalized residue in the first iteration is obtained as 
\begin{equation}
\tilde{\rbf} = \bm{\Omega}\rbf = \bm{\Omega} \Wbf \zbf,
\end{equation}
where $\bm{\Omega} = \bm{\Omega}^{(1)}$ is defined as in (\ref{eq:norm_matrix}). 

  Due to the normalization, each entry $\tilde{r}_{i}$ is distributed as $\Nc(0,1)$ unless $\{i\}$ is a critical set \cite{Abur&Exposito:book}; if $\{i\}$ is a critical set, $\tilde{r}_{i}$ is equal to zero for any $\zbf$.
 
If an attack vector $\abf$ is added, the resulting normalized residue 
is
\begin{equation}\label{eq:attack_on_r_N}
\tilde{\rbf} = \bm{\Omega} \Wbf (\zbf + \abf) = \bm{\Omega} \Wbf\zbf + \bm{\Omega} \Wbf\abf.
\end{equation}
Thus, if $\{i\}$ is not a critical set, $\tilde{r}_{i}$ is distributed as $\Nc((\bm{\Omega} \Wbf\abf)_{i}, 1)$; if $\{i\}$ is critical, $\tilde{r}_{i} = (\bm{\Omega} \Wbf\abf)_{i}$ surely.

Recalling that the absolute normalized residues (\ie, $|\tilde{r}_{i}|$) are the statistics used for identifying the bad data entries, one intuitive heuristic to frame the meters in $\SmscF$ as bad is to make the mean energy of the normalized residues at the framed meters as large as possible.  
Making the framed meters have large normalized residues at the first iteration is of course not a guarantee that their data will be identified as bad and removed in the subsequent iterations.   
Nevertheless, 
 this is a reasonable heuristic to avoid the difficult task of analyzing the dynamic adversarial effect in subsequent iterations.  
Note that the expected energy of the normalized residues at the framed meters is 
\begin{equation}\label{eq:mean_energy_quad}
\mbbE\left[\sum_{i\in\Smsc_{\mbox{\tiny F}}}(\tilde{r}_{i})^{2}\right] = \sum_{i\in\Smsc_{\mbox{\tiny F}}}\mbbE[(\tilde{r}_{i})^{2}] = \sum_{i\in\Smsc_{\mbox{\tiny F}}}(\bm{\Omega} \Wbf\abf)_{i}^{2} + C,
\end{equation}
where $C$ is the number of the meters in $\SmscF$ that do not form a single-element critical set. 
Therefore, maximizing the mean energy of the normalized residues at $\SmscF$ is equivalent to maximizing $\sum_{i\in\Smsc_{\mbox{\tiny F}}}(\bm{\Omega} \Wbf\abf)_{i}^{2}= \|\Rbf_{\Fscr}\bm{\Omega} \Wbf\abf\|_{2}^{2}$ where $\Rbf_{\Fscr}\in\mbbR^{|\Smsc_{\mbox{\tiny F}}|\times m}$ is the row-selection matrix that retains only the rows corresponding to the framed meters.

\subsection{Optimal framing attack via QCQP}\label{subsec:optimization}

The ultimate objective of a framing attack is to gain an ability to perturb the state estimate by an arbitrary degree.  
To this end, a framing attack aims to accomplish two tasks.

The first is to make the bad data processing remove the framed meters such that a covert attack exists in the network with the remaining meters.  
As discussed in Section~\ref{subsec:effect_rN}, we attempt to achieve this goal by maximizing the mean energy of the normalized residues at $\SmscF$, which is equivalent to maximizing $\|\Rbf_{\Fscr}\bm{\Omega} \Wbf \abf\|_{2}^{2}$.    

The second task is to ensure that the attack becomes covert once the framed meters are removed, thereby making the attack as effective as a covert attack (\ie, enable perturbation of the state estimate by an arbitrary degree.)  
Let $\Hbf_{0}$ denote the $m\times n$ measurement matrix obtained from $\Hbf$ by replacing the rows corresponding to the framed meters with zero row vectors. 
Then, the attack becomes covert after the framed meters are removed (\ie, the attack vector lies in the column space of the remaining measurement matrix,) if and ony if $\abf$ is in $\Rmsc(\Hbf_{0})$.  
Therefore, we restrict the attack vector $\abf$ to be not only in the feasible set $\Amsc$ but also in $\Rmsc(\Hbf_{0})$.

Based on the aforementioned intuition, we solve the following optimization to find the optimal \emph{direction} to align the attack vector:
\begin{equation}\label{eq:high_level_des}
\begin{array}{ll}
\max_{\abf} & \|\Rbf_{\Fscr}\bm{\Omega} \Wbf \abf\|_{2}^{2}\\
\text{subj.} & \|\abf\|_{2}^{2} = 1, ~~ \abf\in\Rmsc(\Hbf_{0})\cap\Amsc.
\end{array}
\end{equation}
The optimization (\ref{eq:high_level_des}) gives the optimal direction $\abf^{*}$ of an attack vector that maximizes the mean energy of the normalized residues at $\SmscF$, among the feasible directions that render the attack covert after the framed meters are removed.

To provide a more intuitive description of the feasible set of (\ref{eq:high_level_des}), we introduce 
 the $(m-|\SmscA| - |\SmscF|)\times n$ matrix $\bar{\Hbf}$ obtained from $\Hbf$ by removing the rows corresponding to the adversary meters and the framed meters.  
It can be easily seen that $\abf\in\Rmsc(\Hbf_{0})\cap\Amsc$ if and only if $\abf = \Hbf_{0}\xbf_{0}$ for some $\xbf_{0}\in\Nmsc(\bar{\Hbf})$.  
Therefore, the dimension of $\Rmsc(\Hbf_{0})\cap\Amsc$ is equal to the dimension of $\Nmsc(\bar{\Hbf})$. 
For instance, if $\SmscA\cup\SmscF$ is a critical set, $\bar{\Hbf}$ has rank $n-1$, and its null space has dimension one. 
 Therefore, in this case, $\Rmsc(\Hbf_{0})\cap\Amsc$ is a one-dimensional space, and there is no need to search for the optimal direction. 
On the other hand, if $\SmscA\cup\SmscF$ contains more than one critical sets, the dimension of $\Nmsc(\bar{\Hbf})$ is greater than one, and the optimization (\ref{eq:high_level_des}) searches for the optimal direction in the infinite set of feasible directions.  

Finally, we set an attack vector $\abf$ as $\eta\cdot\abf^{*}$ where $\eta\in\mbbR$ is a parameter that adjusts the direction (\ie, positive or negative depending on the sign of $\eta$) and the magnitude of the resulting perturbation of the state estimate.
The injected attack vector will cause the bad data detector to invoke the bad data identification mechanism for several iterations. 

It is important to point out that removal of all the framed meters is \emph{not} necessary although it does guarantee the successful perturbation of the state estimate.  
Let $N$ denote the total number of iterations in state estimation.  
Then, the state estimate is obtained by finding the linear WLS estimate (\ref{eq:WLS_DC}) to fit $\zbf^{(N)}$ using the columns of $\Hbf^{(N)}$.   
And, as long as $\zbf^{(N)}$ contains a meter data modified by the attacker, 
 the state estimate will be distorted to fit the adversarially modified data.  
In some circumstances, an attack vector designed by (\ref{eq:high_level_des}) might cause removal of only a part of the framed meters and some of the adversary meters.  Nevertheless, such removal may still be able to make some adversary meter data remain in $\zbf^{(N)}$ thereby leading to a successful attack.  



The optimization (\ref{eq:high_level_des}) can be written as a QCQP:
\begin{equation}\label{eq:qcqp_y}
\begin{array}{ll}
\min_{\qbf} & \qbf^{T}\Pbf\qbf\\
\text{subj.} & \qbf^{T}\Qbf\qbf -1 = 0,~~\qbf\in\mbbR^{p},
\end{array}
\end{equation}
where
\begin{equation}
\Pbf \triangleq  - (\Rbf_{\Fscr}\bm{\Omega} \Wbf \Bbf)^{T}(\Rbf_{\Fscr}\bm{\Omega} \Wbf \Bbf),~~ \Qbf \triangleq \Bbf^{T}\Bbf,
\end{equation}
and $\Bbf\in\mbbR^{m\times p}$ is the basis matrix of the $p$-dimensional vector space $\Rmsc(\Hbf_{0})\cap\Amsc$.
Note that the dimension $p$ is nonzero because $\SmscF$ is set such that the feasible set of (\ref{eq:high_level_des}) is nonempty. 
In addition, $\Pbf$ is negative semidefinite, and $\Qbf$ is positive definite since $\Bbf$ has full column rank.
The positive definiteness of $\Qbf$ implies that a solution exists (\ie, the objective function is bounded below.) 

The KKT conditions for (\ref{eq:qcqp_y}) are as follows:
\begin{equation}\label{eq:KKT}
\Pbf\qbf + \lambda(\Qbf\qbf) = 0,~~\qbf^{T}\Qbf\qbf -1 = 0,
\end{equation}
where $\lambda$ is the Lagrange multiplier for the equality constraint.
The optimal solution $\qbf^{*}$ of (\ref{eq:qcqp_y}) is the one that results in the minimum objective function value among all $(\lambda, \qbf)$ pairs satisfying the KKT conditions (\ref{eq:KKT}).  

The KKT conditions (\ref{eq:KKT}) imply that
\begin{equation}
\begin{array}{ll}
\Qbf^{-1}\Pbf\qbf &= \lambda \qbf;\\[2pt]
\qbf^{T} \Pbf\qbf &= \qbf^{T}( -\lambda \Qbf\qbf) = -\lambda\qbf^{T} \Qbf\qbf = -\lambda.
\end{array}
\end{equation}
For any solution $(\lambda, \qbf)$ of (\ref{eq:KKT}), the first equation means that $\lambda$ should be an eigenvalue of $\Qbf^{-1}\Pbf$, and $\qbf$ should be in the corresponding eigenspace.
The second equation means that the objective function value at $\qbf$ is equal to $-\lambda$.
Therefore, we can find an optimal solution $\qbf^{*}$ of (\ref{eq:qcqp_y}) as follows: (i) find the maximum eigenvalue of $\Qbf^{-1}\Pbf$, and (ii) find an eigenvector $\qbf^{*}$ in the corresponding eigenspace that satisfies $(\qbf^{*})^{T}\Qbf\qbf^{*} -1 = 0$.
Once $\qbf^{*}$ is found, an optimal solution $\abf^{*}$ of the original problem (\ref{eq:high_level_des}) is obtained as $\abf^{*} = \Bbf\qbf^{*}$.  

\section{Factor-of-two result}\label{sec:factor_of_two}

In this section, we demonstrate that a framing attack enables the attacker controlling only a half of a critical set of meters to perturb the state estimate by an arbitrary degree.  
Specifically, given a partition $\{\Smsc_{1}, \Smsc_{2}\}$ of a critical set of meters, we present a sufficient condition under which the attacker can control one of $\Smsc_{1}$ or $\Smsc_{2}$ to perturb the state estimate by an arbitrary degree.  
We provide the numerical evidence based on the IEEE benchmark networks that for a critical set associated with a cut, we can find a partition with $|\Smsc_{1}| \simeq |\Smsc_{2}|$ satisfying the sufficient condition. 

\subsection{Estimation of adversarial state estimate perturbation}\label{subsec:estimate_effect}
The exact analysis of how a framing attack would perturb the state estimate is a difficult task due to the iterative nature of state estimation and bad data processing.  However, assuming that meter SNRs are high, we can estimate the effect of a framing attack as follows.  Since SNRs of most practical meters tend to be higher than 46 dB\cite{EPRIwhitepaper}, the high meter SNR assumption is reasonable.

Suppose that the attacker adds an attack vector $\abf$ to $\zbf$, and the iterative state estimation is executed on $\bar{\zbf}$.  The measurement vector at the $k$th iteration is
\begin{equation}
\bar{\zbf}^{(k)} = \Hbf^{(k)}\xbf + \abf^{(k)} +\ebf^{(k)},
\end{equation}
where $\Hbf^{(k)}$, $\abf^{(k)}$, and $\ebf^{(k)}$ are obtained from $\Hbf$, $\abf$, and $\ebf$ by removing the $(k-1)$ rows corresponding to the meters identified as bad until the $(k-1)$st iteration.
The state estimate $\hat{\xbf}^{(k)}$ at the $k$th iteration is
\begin{equation}
\begin{array}{l}
[(\Hbf^{(k)})^{T}(\bm{\Sigma}^{(k)})^{-1}\Hbf^{(k)}]^{-1}(\Hbf^{(k)})^{T}(\bm{\Sigma}^{(k)})^{-1}\bar{\zbf}^{(k)}\\
= \xbf + [(\Hbf^{(k)})^{T}(\bm{\Sigma}^{(k)})^{-1}\Hbf^{(k)}]^{-1}(\Hbf^{(k)})^{T}(\bm{\Sigma}^{(k)})^{-1}\\
\hfill \cdot(\abf^{(k)} +\ebf^{(k)}).
\end{array}
\end{equation}
Hence, the state estimate error at the $k$th iteration is
\begin{equation}\label{eq:state_perturb}
\begin{array}{l}
\hat{\xbf}^{(k)} - \xbf = [(\Hbf^{(k)})^{T}(\bm{\Sigma}^{(k)})^{-1}\Hbf^{(k)}]^{-1}(\Hbf^{(k)})^{T}(\bm{\Sigma}^{(k)})^{-1}\\
\hfill\cdot(\abf^{(k)} +\ebf^{(k)}).
\end{array}
\end{equation}
In addition, the residue vector is
\begin{equation}\label{eq:residue_k}
\begin{array}{ll}
\rbf^{(k)} &= \Wbf^{(k)}\bar{\zbf}^{(k)}= \Wbf^{(k)}(\Hbf^{(k)}\xbf + \abf^{(k)} + \ebf^{(k)})\\[3pt]
& = \Wbf^{(k)}(\abf^{(k)} + \ebf^{(k)}).
\end{array}
\end{equation}
From (\ref{eq:state_perturb}) and (\ref{eq:residue_k}), we can see that both the state estimate error and the residue vector do not depend on the actual state $\xbf$.
Considering that bad data detection and identification at each iteration exclusively rely on the residue vector, the observation from (\ref{eq:state_perturb}) and (\ref{eq:residue_k}) implies that if we aim to analyze how much the attack would {perturb} the final state estimate, \ie, $\hat{\xbf}^{(N)} - \xbf$, where $N$ denotes the total number of iterations, we can simply work with $\abf +\ebf$ by assuming that $\xbf$ is equal to $\textbf{0}$. 

Furthermore, if meter SNRs are significantly large (\ie, $\sigma^{2}\ll 1$), we can estimate the resulting perturbation of the state estimate by running the \emph{noiseless} version of the iterative state estimation on the attack vector $\abf$ and checking the resulting $\hat{\xbf}^{(N)}$.  The {noiseless} version means the algorithm which the iterative state estimation converges to as $\sigma^{2}$ decays to 0.  Specifically, $\bm{\Sigma}$ is replaced\footnote{Note that state estimation and bad data identification are not affected by the value of $\sigma^{2}$. Because, $\sigma^{2}$ gets cancelled out in the state estimate expression (\ref{eq:WLS_DC}), and bad data identification relies on the relative order of the normalized residue magnitudes, which are not affected by the value of $\sigma^{2}$.  Only bad data detection is affected by the decaying $\sigma^{2}$.} by $\bar{\bm{\Sigma}}$, and at the $k$th iteration, the bad data detector declares presence of bad data if and only if $(\rbf^{(k)})^{T}(\bar{\bm{\Sigma}}^{(k)})^{-1}\rbf^{(k)} > 0$ (\ie, the data are declared to be good if and only if state estimation results in a zero residue vector.)

\subsection{Factor-of-two theorem for critical sets}

Suppose that $\{\Smsc_{1}, \Smsc_{2}\}$ is a partition of a critical set, and let $\bar{\Hbf}$ denote the measurement matrix after removing the meters in $\Smsc_{1}\cup\Smsc_{2}$ from the network.  Since $\Smsc_{1}\cup\Smsc_{2}$ is a critical set, $\bar{\Hbf}$ has rank $n-1$, and the dimension of its null space is one.  Let $\Delta\xbf$ denote a unit basis vector of the null space of $\bar{\Hbf}$.
Recalling the discussion in Section~\ref{subsec:optimization}, 
if $\Smsc_{1}$ is the set of adversary meters, and $\Smsc_{2}$ is the set of framed meters, then the framing attack aligns the attack vector along $\Hbf_{1}\Delta\xbf$, where $\Hbf_{1}$ is the $m\times n$ matrix obtained from $\Hbf$ by replacing the rows corresponding to the meters in $\Smsc_{2}$ with zero row vectors ($\Hbf_{2}$ is defined in the same way by replacing the rows of $\Hbf$ corresponding to $\Smsc_{1}$ with zero row vectors.) 

The following theorem provides a sufficient condition that guarantees that a framing attack can use one of $\Smsc_{1}$ and $\Smsc_{2}$ to perturb the state estimate by an arbitrary degree under the high SNR setting.
The condition is based on the result of executing the analysis described in Section~\ref{subsec:estimate_effect}.

\slisp
\begin{theorem}\label{thm:perturb}
Suppose that if we run the noiseless version of the iterative state estimation on $\Hbf_{1}\Delta\xbf$, then there exists a unique state $\ybf\in\mbbR^{n}$ such that the final state estimate is always equal to $\ybf$ (\ie, $\hat{\xbf}^{(N)} = \ybf$) regardless of whatever decisions are made under tie\footnote{It is possible that a \emph{tie} may occur in bad data identification at some iteration: \ie, the largest absolute normalized residue is assumed by more than one meters.  In a tie situation, we assume that bad data identification chooses an arbitrary meter with the largest absolute normalized residue.} situations in bad data identification.  Under this condition, the following statements hold for any true state $\xbf\in\mbbR^{n}$:

(1) 
 If a framing attack using $\Smsc_{1}$ and $\Smsc_{2}$  as $\SmscA$ and $\SmscF$ respectively (\ie, $\abf = \eta\cdot \Hbf_{1}\Delta\xbf$ where $\eta\in\mbbR$ is a scaling factor) is launched, then
\begin{equation}\label{eq:thm_event}
\lim_{\sigma^{2}\rightarrow 0}\Pr(\bar{\zbf}^{(N)} = \Hbf^{(N)}(\xbf + \eta\cdot\ybf) + \ebf^{(N)}) = 1,
\end{equation}
where $N$ is the random variable representing the total number of iterations in the iterative state estimation.\slisp

(2) 
 If a framing attack using $\Smsc_{2}$ and $\Smsc_{1}$  as $\SmscA$ and $\SmscF$ respectively (\ie, $\abf = \eta\cdot \Hbf_{2}\Delta\xbf$) is launched, 
\begin{equation}
\lim_{\sigma^{2}\rightarrow 0}\Pr(\bar{\zbf}^{(N)} = \Hbf^{(N)}(\xbf +\eta\cdot(\Delta\xbf - \ybf))  + \ebf^{(N)}) = 1.
\end{equation}

\end{theorem}

\emph{Proof:} See Appendix.
\endproof
\slisp

\emph{Remark:} If no tie occurs in bad data identification, the condition of Theorem~\ref{thm:perturb} is naturally satisfied.  Note that a tie in bad data identification is a rare event, so the condition is likely to hold for a general partition of a critical set.

The event $\{\bar{\zbf}^{(N)} = \Hbf^{(N)}(\xbf + \eta\cdot\ybf) + \ebf^{(N)}\}$ in (\ref{eq:thm_event}) means that the remaining measurements at the final iteration of state estimation appear to be normal measurements from the perturbed state $\xbf + \eta\cdot\ybf$. 
Theorem~\ref{thm:perturb} implies that if the condition is met, then at least one of $\Smsc_{1}$ and $\Smsc_{2}$ can be used as the set of adversary meters by a framing attack to perturb the state estimate by an arbitrary degree, because $\ybf$ and $\Delta\xbf - \ybf$ cannot be simultaneously $\textbf{0}$.  
Especially, if the condition holds for a partition with $|\Smsc_{1}| = |\Smsc_{2}|$, then the adversary controlling only a half of the critical set can perturb the state estimate by an arbitrary degree.

\begin{figure}[t!]
\centering
\psfrag{z}[c]{ $\zbf$ }
\psfrag{zb}[c]{ $\bar{\zbf}$ }
\psfrag{g}[c]{ $\Gmsc$}
\psfrag{gb}[c]{ ${\Gmsc}$}
\psfrag{x}[c]{ $\hat{\xbf}$ }
\psfrag{o}[l]{ $(\hat{\xbf},\Gmsc)$}
\psfrag{ob}[l]{$\hat{\xbf}$}
\psfrag{n}[c]{ $\sbf$}
\includegraphics[width=.5\textwidth]{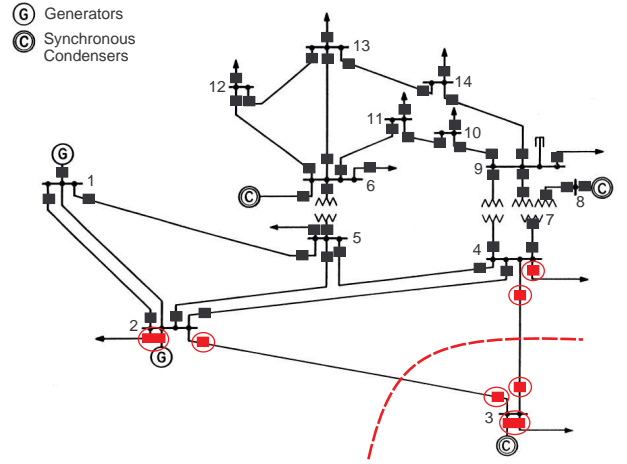}
\caption{IEEE 14-bus network:  the rectangles on lines and buses represent line flow meters and bus injection meters respectively.  The line meter on the line $\{i, j\}$, that is closer to $i$, measures the power flow from $i$ to $j$.  The red dashed line describes a cut, and the circled meters are the meters associated with the cut.}
\label{fig:cut_14}
\end{figure}

One important question is whether a partition $\{\Smsc_{1}, \Smsc_{2}\}$ with $|\Smsc_{1}| \simeq |\Smsc_{2}|$ that satisfies the condition of Theorem~\ref{thm:perturb} can be found in general.
To answer this question, we investigated critical sets \emph{associated with cuts}\footnote{A cut of an undirected graph $(\Vmsc,\Emsc)$ is defined as a partition $\{\Vmsc_{1}, \Vmsc_{2}\}$ of $\Vmsc$ consisting of two nonempty subsets, and the associated cut-set is the subset of lines connecting two vertices in different partitions.} in the IEEE 14-bus network and the IEEE 118-bus network, where every bus has an injection meter and every line has line meters for both directions.  
The spanning tree observability criterion in \cite{Krumpholz&Clements&Davis:80PAS} implies that the set $\Smsc$ of the meters associated with a cut (\ie, the set of all line meters on the cut-set lines and all injection meters at the endpoints of the cut-set lines) is a critical set if the cut decomposes the network topology into two connected graphs.  For instance, the cut in Fig.~\ref{fig:cut_14} disconnects the bus 3 from the rest of the network, and $\{\{2,3\}, \{3,4\}\}$ is the associated cut-set.  The set of circled red meters is the critical set associated with the cut. 

We executed 20,000 runs of the random contraction algorithm by Karger and Stein \cite{Karger&Stein:1996JACM}---a randomized algorithm for finding a cut---and found 118 cuts in the 14-bus network and 290 cuts in the 118-bus network.
For each cut, we found a partition $\{\Smsc_{1}, \Smsc_{2}\}$ of the critical set $\Smsc$ associated with the cut such that $|\Smsc_{1}|\simeq \frac{|\Smsc|}{2}$: $\Smsc_{1}$ consists of the line meters (both directions) on a subset of cut-set lines such that $\left||\Smsc_{1}| - \frac{|\Smsc|}{2}\right|\leq 1$, and $\Smsc_{2}$ is set to be $\Smsc\setminus\Smsc_{1}$.  
In both networks, for every cut we considered, the partition constructed in the aforementioned manner satisfied the condition of Theorem~\ref{thm:perturb}; this suggests that controlling about a half of a critical set associated with a cut, a framing attack may perturb the state estimate by an arbitrary degree\footnote{The average size of the critical sets we considered is 15.7 for the 14-bus network and 12.7 for the 118-bus network.}.   

\section{Numerical results}\label{sec:numerical_baddata}

We tested the performance of framing attacks with the IEEE 14-bus network and the IEEE 118-bus network using the AC model and the nonlinear iterative state estimation described in Section~\ref{sec:SE}. 
The simulation results demonstrate the efficacy of framing attacks under the real-world power system setting.
Because the ultimate goal of the attack is to perturb the state estimate, we measure the mean $l_2$-norm of the resulting state estimate error:
$$\mbbE[\|\hat{\xbf} - \xbf\|_{2}],$$
where $\hat{\xbf}$ is the state estimate, and $\xbf$ is the true state.

\subsection{Simulation setting}
For each test network, we chose representative attack scenarios (\ie, $\SmscA$ and $\SmscF$) and tested the performance of framing attacks.  
For each case, we ran Monte Carlo simulations to evaluate the mean state estimate error.  In each Monte Carlo run, the true state $\xbf$ was generated by a multivariate Gaussian distribution with small variances\footnote{The standard deviation of each phase angle is set to be 1.15 degree.  The standard deviation of each voltage magnitude is set to be 0.01 p.u.}.  Its mean was set as the operating state given by the IEEE data\cite{IEEEParameter}, which is far from the nominal state used to obtain the DC model.  
Based on the state $\xbf$, noisy measurements were generated by the AC model $(\ie, h(\xbf) + \ebf)$.  
The attack vector was constructed based on the DC measurement matrix $\Hbf$ as described in Section~\ref{sec:frame_attack}. 
Once constructed, the attack vector was added to the noisy measurements, and the iterative state estimation\footnote{The false alarm rate of the bad data detector is set to be 0.04 throughout all the simulations.} was executed on the corrupted measurements.  After the iterative state estimation finished, we measured $\|\hat{\xbf}^{(N)} - \xbf\|_{2}$.

Note that the design of a framing attack was studied using the DC model which has only the real part of the measurements.
For the simulations, we designed an attack vector based on the DC model, and the attack modified only the corresponding real part of the measurements.
Considering the linear decoupled model (see Chapter 2.7 in \cite{Abur&Exposito:book}), such an attack is expected to modify primarily the bus voltage phase angles and have little effect on the bus voltage magnitudes.
Hence, in interpreting the results, we focus on the perturbation of the phase angle part of the state estimate.

For comparison, we also executed the conservative scheme in \cite{Kosut11}, which aims to perturb the state estimate by the maximum degree while avoiding detection by the bad data detector.  
In the conservative scheme, the attack vector was designed as a solution to
\begin{equation}
\begin{array}{ll}
\text{max}_{\abf\in\Amsc}& \|(\Hbf^{T}\bm{\Sigma}^{-1}\Hbf)^{-1}\Hbf^{T}\bm{\Sigma}^{-1}\abf\|^{2}_{2}\\
\text{subj.}&\rbf^{T}\bm{\Sigma}^{-1}\rbf \leq \tau,
\end{array}
\end{equation}
where the constraint guarantees that the alarm is not raised at all, and the objective function is the resulting perturbation of the state estimate due to the attack vector.  

\subsection{Simulation results with 14-bus network}

We first tested the case where the adversary can control only a half of a critical set.
Specifically, we considered the adversary who can control $(2,3)$, $(3,4)$, and $(4,3)$: $(i,j)$ denotes the line meter for the power flow from $i$ to $j$, and $(i)$ denotes the injection meter at bus $i$.
The framed meters were set to be $(3,2)$, $(2)$, $(3)$, and $(4)$ such that the set of adversary meters and framed meters is the critical set associated with the cut in Fig.~\ref{fig:cut_14}.
We tested framing attacks with three different attack magnitudes: $\|\abf\|_{1}$ is $1\%$, $2\%$, or $3\%$ of $\|\zbf\|_{1}$.

\begin{figure}[t!]
\centering
\includegraphics[width=.48\textwidth]{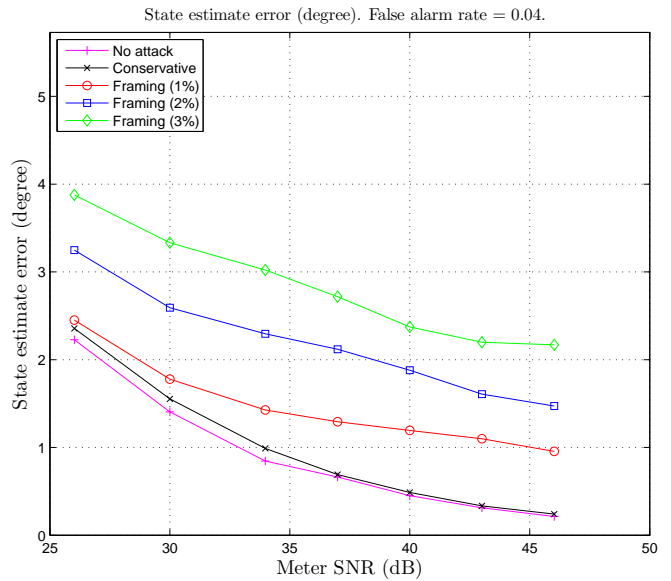}
\caption{The 14-bus network: 1,000 Monte Carlo runs. The adversary meters are $(2,3)$, $(3,4)$, and $(4,3)$, and the framed meters are $(2)$, $(3)$, $(4)$, $(3,2)$.}
\label{fig:23_34_43_AC_abs}
\end{figure}

Fig.~\ref{fig:23_34_43_AC_abs} shows the resulting state estimate error versus the meter SNR.
The meter SNR ranges from 26 dB to 46 dB (equivalently, the noise-to-signal amplitude ratio ranges from 5$\%$ to 0.5$\%$.)   
The normal state estimate error and the state estimate error under the conservative attack are very close, and both decay to zero as the SNR increases.
However, the state estimate errors in the presence of framing attacks converge to constants as the SNR increases, and the constants are proportional to the attack magnitudes especially in the high SNR region.    
The result implies that a framing attack can adjust the state estimate perturbation by choosing a proper attack magnitude.  

\begin{figure}[t!]
\centering
\includegraphics[width=.48\textwidth]{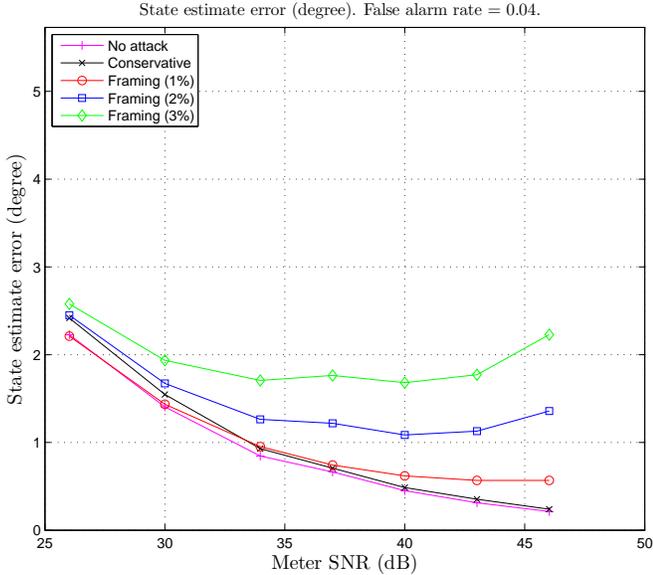}
\caption{The 14-bus network: 1,000 Monte Carlo runs.  The adversary meters are $(2,3)$, $(3,4)$, $(4,3)$, $(6,12)$, $(12,6)$, and $(12,13)$, and the framed meters are $(2)$, $(3)$, $(4)$, $(3,2)$, $(6)$, $(12)$, $(13)$, and $(13,12)$.}
\label{fig:612_126_1213_23_34_43_AC}
\end{figure}

Second, we demonstrate that a framing attack may perturb the state estimate in various directions depending on the choice of the set of framed meters. 
We considered the case that the adversary controls $(2,3)$, $(3,4)$, $(4,3)$, $(6,12)$, $(12,6)$, and $(12,13)$.  Note that the adversary still cannot control any critical set, and thus a covert attack is infeasible.
The framing attack with $\SmscF$ equal to any of the following three sets successfully perturbed the state estimate: (i) $(2)$, $(3)$, $(4)$, $(3,2)$, $(6)$, $(12)$, $(13)$, and $(13,12)$; (ii) $(2)$, $(3)$, $(4)$, and $(3,2)$; (iii) $(6)$, $(12)$, $(13)$, and $(13,12)$.  
For instance, Fig.~\ref{fig:612_126_1213_23_34_43_AC} shows the state estimate error versus the meter SNR for the first set.
While the three sets all resulted in successful state estimate perturbation, each resulted in a different direction of perturbation.
For each $\SmscF$, Table~\ref{table:direction} shows the three buses, whose phase angle estimates were most significantly perturbed, and the mean perturbation of their phase angle estimates; positive perturbation means overestimation, and negative perturbation means underestimation.  
The table demonstrates that the adversary controlling a large number of meters may adjust the direction of perturbation by setting $\SmscF$ properly.   Note that the impact of a framing attack with a specific $\SmscF$ can be estimated based on the analysis in Section~\ref{subsec:estimate_effect}.

\begin{table}[!t]
\caption{The three buses whose phase angles are most significantly perturbed by each attack:  1,000 Monte Carlo runs, SNR = 46dB.}
\renewcommand{\arraystretch}{1.1}
\setlength{\tabcolsep}{0.03in}
\begin{center}
\begin{tabular}{|c|c|c|}\hline
\begin{tabular}{c}
$(2)$, $(3)$, $(4)$, $(3,2)$,\\
 $(6)$, $(12)$, $(13)$, $(13,12)$
\end{tabular} & \begin{tabular}{c}
$(2)$, $(3)$\\ $(4)$, $(3,2)$
\end{tabular} & \begin{tabular}{c}
 $(6)$, $(12)$,\\
  $(13)$, $(13,12)$
\end{tabular} \\ \hline
\begin{tabular}{l}
1) bus 12: $2.075^{\circ}$ \\
2) bus 3: $0.272^{\circ}$\\
3) bus 14: $-0.180^{\circ}$
\end{tabular}
&
\begin{tabular}{l}
1) bus 3: $-2.183^{\circ}$ \\
2) bus 14: $0.182^{\circ}$\\
3) bus 9: $0.168^{\circ}$
\end{tabular}
& 
\begin{tabular}{l}
1) bus 12: $2.878^{\circ}$ \\
2) bus 14: $0.005^{\circ}$\\
3) bus 9: $0.004^{\circ}$
\end{tabular}
\\ \hline
\end{tabular}
\label{table:direction}
\end{center}
\end{table}

\subsection{Simulation results with 118-bus network}

\begin{figure}[t!]
\centering
\includegraphics[width=.48\textwidth]{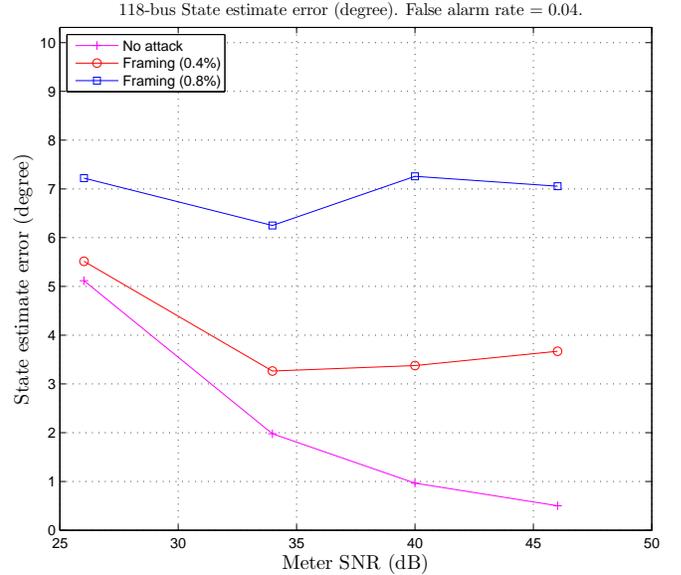}
\caption{The 118-bus network: 250 Monte Carlo runs.  The adversary meters are $(20,21)$, $(21,20)$, and $(21,22)$, and the framed meters are $(20)$, $(21)$, $(22)$, and $(22,21)$.}
\label{fig:118_AC}
\end{figure}

Through the simulations with the 118-bus network, we aim to demonstrate the effect of framing attacks on a larger network.  
We considered the scenario where the adversary controls $(20,21)$, $(21,20)$, and $(21,22)$, and the framed meters are $(20)$, $(21)$, $(22)$, and $(22,21)$; \ie, the set of the adversary meters and the framed meters is the critical set associated with the cut isolating the bus 21 from the rest of the network.  
Fig.~\ref{fig:118_AC} shows the state estimate errors under the non-attack scenario and the framing attacks with different attack magnitudes.
The results imply that the framing attacks successfully perturbed the state estimate in a large network.

\section{Conclusions}\label{sec:conclusion_frame}

We have presented a data framing attack on power system state estimation.  
Controlling only a half of a critical set, the data framing attack can perturb the state estimate by an arbitrary degree. 
A theoretical justification was provided, and numerical experiments demonstrated the efficacy of the framing attack.

Our results indicate that most known countermeasures, that are aimed at merely preventing covert state attacks, are not sufficient for protection against attacks on state estimation.  The proposed framing attacks can successfully perturb the state estimate even when those countermeasures are employed.  
In designing a countermeasure, the possibility of a framing attack needs to be taken into account.

An important direction for future work is to design a mechanism that can nullify the attack impact once the presence of an attack is detected.  
In particular, we need a robust state estimation mechanism that can produce an unbiased state estimate with reasonable accuracy even when some data entries are untrustworthy due to the attacker's modification.


\appendix[Proof of Theorem~\ref{thm:perturb}]

Let $\Smsc$ denote the set of sequences of meter removals that can possibly happen when the noiseless version of the iterative state estimation is executed on $\Hbf_{1}\Delta\xbf$.  In other words, $(a_{1},\ldots,a_{M})\in\Smsc$ if and only if some decisions under tie situations may result in the removal of the meters $\{a_{1},\ldots,a_{M}\}$ in the order of $a_{1},\ldots, a_{M}$.  The cardinality of $\Smsc$ can be greater than $1$ since different decisions under tie situations may result in different sequences of meter removals.


For any sequence $(a_{1},\ldots,a_{M})\in\Smsc$, the existence of such $\ybf$---as described in the condition---implies that
if all the meters in the sequence are removed, the remaining part of $\Hbf_{1}\Delta\xbf$, denoted by $\Hbf_{1}^{(M)}\Delta\xbf$, is equal to $\Hbf^{(M)}\ybf$, where $\Hbf_{1}^{(M)}$ and $\Hbf^{(M)}$ are obtained from $\Hbf_{1}$ and $\Hbf$ respectively, by removing the rows corresponding to all meters in the sequence.

Now, consider running the iterative state estimation on $\Hbf\xbf + \Hbf_{1}\Delta\xbf + \ebf$, which is the resulting measurement vector when the framing attack is launched with $\SmscA$ and $\SmscF$ equal to $\Smsc_{1}$ and $\Smsc_{2}$ respectively.
The equation (\ref{eq:residue_k}) implies that the residue vector in each iteration only depends on $\Hbf_{1}\Delta\xbf + \ebf$. 
In addition, as $\sigma^{2}$ decreases to zero, the results of bad data detection and identification heavily depend on $\Hbf_{1}\Delta\xbf$, and thus the sequence of meter removals becomes highly likely to be in $\Smsc$.  Formally, 
\begin{equation}
\lim_{\sigma^{2}\rightarrow 0}\Pr((a_{1},\ldots,a_{N})\in\Smsc) = 1,
\end{equation}
where $(a_{1},\ldots,a_{N})$ is a sequence of meters removed by the bad data identification rule.  
Let $\Hbf^{(N)}$ and $\ebf^{(N)}$ denote the matrix and vector obtained from $\Hbf$ and $\ebf$ respectively by removing the rows corresponding to $\{a_{1},\ldots,a_{N}\}$.  Note that $N$, $(a_{1},\ldots,a_{N})$, $\Hbf^{(N)}$, and $\ebf^{(N)}$ are all random.

The event $\{(a_{1},\ldots,a_{N})\in\Smsc\}$ implies that $\Hbf_{1}^{(N)}\Delta\xbf = \Hbf^{(N)}\ybf$, and thus
\begin{equation}
\bar{\zbf}^{(N)} = (\Hbf^{(N)}\xbf + \ebf^{(N)}) + \Hbf^{(N)}_{1}\Delta\xbf = (\Hbf^{(N)}\xbf + \ebf^{(N)}) + \Hbf^{(N)}\ybf.
\end{equation}
Therefore,
\begin{equation}\label{eq:H1_prob}
\lim_{\sigma^{2}\rightarrow 0}\Pr(\bar{\zbf}^{(N)} = (\Hbf^{(N)}\xbf + \ebf^{(N)}) + \Hbf^{(N)}\ybf) = 1.
\end{equation}
When the attack vector $\Hbf_{1}\Delta\xbf$ is replaced with $\eta\cdot\Hbf_{1}\Delta\xbf$, we can use the exactly same logic to derive the following.  
\begin{equation}
\lim_{\sigma^{2}\rightarrow 0}\Pr(\bar{\zbf}^{(N)} = (\Hbf^{(N)}\xbf + \ebf^{(N)}) + \eta\cdot\Hbf^{(N)}\ybf) = 1.
\end{equation}

Now, consider running the iterative state estimation over $\Hbf\xbf + \Hbf_{2}\Delta\xbf + \ebf$; this is the case when the framing attack is launched with $\SmscA$ and $\SmscF$ equal to $\Smsc_{2}$ and $\Smsc_{1}$ respectively.  
First, note that 
\begin{equation}
\Hbf\Delta\xbf = \Hbf_{1}\Delta\xbf + \Hbf_{2}\Delta\xbf.
\end{equation}
Therefore, $\Hbf\xbf + \Hbf_{2}\Delta\xbf + \ebf$ is equivalent to $\Hbf(\xbf + \Delta\xbf) - \Hbf_{1}\Delta\xbf + \ebf$.

Suppose we run the noiseless version of the iterative state estimation on $-\Hbf_{1}\Delta\xbf$.
The set of sequences of meter removals that can possibly happen is equivalent to $\Smsc$, because the sign change only flips the signs of residue entries; it does not affect their absolute values, which are the statistics used for detection and identification of bad data entries.  
  Furthermore, it can be easily seen that the final state estimate is always equal to $-\ybf$ regardless of whatever decisions are made under the tie situations.

The above paragraph implies that we can analyze the result of running the iterative state estimation on $\Hbf(\xbf + \Delta\xbf) - \Hbf_{1}\Delta\xbf + \ebf$ by following the same procedure of deriving (\ref{eq:H1_prob}).  In particular, one can easily derive the following:
\begin{equation}\label{eq:H2_prob}
\lim_{\sigma^{2}\rightarrow 0}\Pr(\bar{\zbf}^{(N)} = \Hbf^{(N)}(\xbf + \Delta\xbf) + \ebf^{(N)} + \Hbf^{(N)}(-\ybf)) = 1,
\end{equation}
or equivalently,
\begin{equation}\label{eq:H2_prob_2}
\lim_{\sigma^{2}\rightarrow 0}\Pr(\bar{\zbf}^{(N)} = \Hbf^{(N)}\xbf + \ebf^{(N)} + \Hbf^{(N)}(\Delta\xbf -\ybf)) = 1.
\end{equation}
When the attack vector $\Hbf_{2}\Delta\xbf$ is scaled by $\eta$ (\ie, $\abf = \eta\cdot\Hbf_{2}\Delta\xbf$,) we can derive the following in a similar manner:
\begin{equation}\label{eq:H2_prob_3}
\lim_{\sigma^{2}\rightarrow 0}\Pr(\bar{\zbf}^{(N)} = \Hbf^{(N)}\xbf + \ebf^{(N)} + \eta\cdot\Hbf^{(N)}(\Delta\xbf -\ybf)) = 1.
\end{equation}
Therefore, the proof is complete. \endproof


\end{document}